# Reconstruction of Antarctic sea ice thickness from sparse satellite laser altimetry data using a partial convolutional neural network

Ziqi Ma[1], Qinghua Yang[1], Yue Xu[2], Wen Shi[3], Xiaoran Dong[1], Qian Shi[1], Hao Luo[1], Jiping Liu[1], Petteri Uotila[4], Yafei Nie[1]

[1]School of Atmospheric Sciences, Sun Yat-sen University and Southern Marine Science and Engineering Guangdong Laboratory (Zhuhai), Zhuhai, 519082, China
[2]School of Earth and Space Sciences, Peking University, Beijing, China
[3]Ministry of Education Key Laboratory for Earth System Modeling, Department of Earth System Science, Tsinghua University, Beijing, 100084, China
[4]Institute for Atmospheric and Earth System Research/ Physics, Faculty of Science, University of Helsinki, Helsinki, Finland

Correspondence to: Yafei Nie (nieyafei@sml-zhuhai.cn)

## Abstract

The persistent lack of spatially complete Antarctic sea ice thickness (SIT) data at sub-monthly resolution has fundamentally constrained the quantitative understanding of large-scale sea ice mass balance processes. In this study, a pan-Antarctic SIT dataset at 5-day and 12.5 km resolution was developed based on sparse Ice, Cloud and Land Elevation Satellite (ICESat; 2003–2009) and ICESat-2 (2018–2024) along-track laser altimetry SIT retrievals using a deep learning approach. The reconstructed SIT was quantitatively validated against independent upward-looking sonar (ULS) observations and showed higher accuracy than the other four satellite-derived and reanalyzed Antarctic SIT datasets. The temporal evolution of the reconstructed SIT was further validated by ULS and ICESat-2 observations. Consistent seasonal cycles and intra-seasonal tendencies across these datasets confirm the reconstruction's reliability. Beyond advancing the mechanistic understanding of Antarctic sea ice variability and climate linkages, this reconstruction dataset's near-real-time updating capability offers operational value for monitoring and forecasting the Antarctic sea ice state.

## Background & Summary

Antarctic sea ice is a vital part of the global climate system. Satellite observations since 1979 have revealed an overall increase in Antarctic sea ice extent until 2016, when an unprecedented dip initiated a new state with frequent record lows[1,2]. More importantly, sea ice volume also exhibited a synchronous precipitous decline in 2016[3,4], with sea ice thickness (SIT) variability being the key driver of volume variability[4]. Yet SIT has remained an observational challenge with spatio-temporal resolution trade-offs[5,6] and insufficient coverage[7,8]. These limitations impede our understanding of sea ice mass balance and evolutionary processes[5,9-11].

Pan-Antarctic SIT observations are primarily obtained through satellite remote sensing. The Soil Moisture and Ocean Salinity (SMOS) satellite measures SIT using a Microwave Imaging Radiometer at L-band, achieving full daily coverage of the Southern Ocean[8]. However, since the brightness temperature saturates with increasing SIT, this method is only effective for detecting thin ice below 1 m during the freezing season. Radar altimeters[4,12,13] (e.g., CryoSat-2 and EnviSat) estimate SIT by measuring the ice freeboard, which is the height of the snow-ice interface above sea level. Combined with snow depth and density data, this freeboard is converted to SIT using the hydrostatic equilibrium principle. However, the retrieval faces significant uncertainties due to challenges in accurately identifying the snow-ice interface[14], which is often flooded with seawater[4]. Laser altimeters[5,15-20], such as NASA's Ice, Cloud, and land Elevation Satellite (ICESat) and its 2$^{nd}$-generation (ICESat-2), mitigate snow metamorphism errors by measuring snow surface-to-sea surface distance (i.e., the total freeboard), currently providing the most reliable satellite-derived SIT estimates[19,21]. However, due to their narrow swaths, these altimeters provide only sparse along-track measurements daily. To obtain spatially complete Antarctic SIT fields, the standard approach aggregates all along-track observations over a month (or season) into a composite map and interpolates the gaps between the tracks[5]. This approach compromises temporal resolution and reduces potential accuracy by assuming that daily measurements represent monthly averages[6].

While in the Southern Ocean, assimilating SIT observations into ice-ocean coupled models facilitates the reconstruction of spatio-temporally complete SIT fields, to date, successful implementation has only been achieved with SMOS thin-ice retrievals[22]. The assimilation of along-track ICESat/ICESat-2 SIT observations remains technically unrealized. Existing ocean reanalyses covering the Southern Ocean that assimilate sea ice concentration (SIC), e.g., the NCEP Climate Forecast System Reanalysis[23] (CFSR), the CMCC Global Ocean Physical Reanalysis System[24] (C-GLORS), the CMEMS global ocean 1/12° reanalysis product[25] (GLORYS12v1), the Global Ice-Ocean Modeling and Assimilation System[26] (GIOMAS), the ECMWF Ocean ReAnalysis System 5[27] (ORAS5), the Southern Ocean State Estimate[28] (SOSE) and the Data Assimilation System for the Southern Ocean (DASSO)[3,22,29], exhibit significant SIT estimation biases[30-32].

Deep learning has emerged as a paradigm-shifting approach for climate data reconstruction, with partial convolutional neural network[33] (PCNN) demonstrating unique advantages in data reconstruction[34-36]. Unlike conventional interpolation methods, such as Kriging and infilling based on principal component analysis, PCNN effectively preserves the continuity of physical fields by adaptive feature extractions[34]. In this study, we aim to reconstruct a 5-day moving averaged pan-Antarctic SIT dataset based on sparse along-track SIT observations from ICESat and ICESat-2 satellites. This dataset will enhance the monitoring of Antarctic sea ice mass balance, support model validation, and facilitate studies on atmosphere-sea ice-ocean interactions and sea ice prediction.

## Methods

The data used in this study are described below and listed in Tables 1 and 2.

**Sea ice thickness and concentration observations**

ICESat and ICESat-2 have provided high-precision measurements of surface elevation and freeboard. ICESat[37], operated from 2003 to 2009, was equipped with the Geoscience Laser Altimeter System (GLAS), which utilized a near-infrared (1,064 nm) laser to measure surface elevation with an accuracy of 3–4 cm over flat sea ice surfaces[38]. Total freeboard was derived from elevation data from its 70 m footprint and 172 m along-track sampling using the Lowest Level Elevation method[15,18,39,40], achieving an accuracy of 2 cm. ICESat-2, launched in October 2018, is equipped with the Advanced Topographic Laser Altimetry System (ATLAS), which achieves an elevation accuracy of 2 cm at an 11 m spatial resolution[41,42]. The ATLAS operates at a 10-kHz pulse rate, generating six photon-counting beams (0.7 m interpulse spacing, ~17 m footprints) grouped into three pairs separated by 3.3 km, each containing a strong and a weak beam. ICESat-2 freeboard estimates are available in the ATL10 product[43], which provides averaged measurements over 10 km segments along the track.

We estimate SIT using the improved one-layer method[19] (OLMi), which utilizes total freeboard from ICESat and ICESat-2 altimetry data. Inheriting the conventional single-layer assumption[15,16], OLMi represents the combined sea ice and snow as an equivalent sea ice layer with reduced density. The final SIT is calculated by subtracting the snow depth, which is estimated based on an empirical ice-snow ratio, from this equivalent layer. Compared to previous single-layer approaches, OLMi reduces the root-mean-square error (RMSE) of SIT estimates by 40–60%[19]. Notably, despite differences in sensor configurations between ICESat and ICESat-2, OLMi ensures consistency in SIT retrievals across missions.

The daily 12.5 km gridded SIC dataset is derived from the brightness temperature of the Advanced Microwave Scanning Radiometer for EOS (AMSR-E; June 2002 to October 2011) and AMSR2[44] (July 2012 to present) using the Enhanced NASA Team algorithm (NT2). The Japan Aerospace Exploration Agency (JAXA) has cross-calibrated the data to ensure temporal consistency, with particular efforts to address the 11-month data gap between the two missions.

Table 1. Summary of data and the periods used in this study. The detailed period of ICESat data used in this study is shown in Table 2.

| | Data name | Period | Temporal resolution | Spatial resolution |
|---|---|---|---|---|
| Input data for reconstruction | ICESat | 2003–2009 (Interruption exists) | - | - |
| | ICESat-2 | 2018–2024 | - | - |
| | AMSR-E | 2003–2009 | Daily | 12.5km |
| | AMSR2 | 2018–2024 | Daily | 12.5km |
| Reanalysis SIT for training PCNN | GLORYS12v1 | 1993–2020 | Daily | 1/12° |
| | C-GLORSv7 | 1989–2019 | Daily | 0.25°×0.1° |
| SIT for comparison and validation | LEGOS | 2003–2009 | Monthly | 12.5km |
| | SICCI | 2003–2009 | Monthly | 50km |
| | ULS | 2003–2009 | 2~15min | - |

**Ocean reanalysis data for training PCNN**

The data-driven reconstruction framework relies on relatively reliable training datasets. Although current ocean-sea ice reanalysis products exhibit substantial biases in SIT estimation, GLORYS12v1 and C-GLORS have been demonstrated to have comparable SIT spatial distributions to the ICESat-corrected benchmark[45]. Given that the PCNN is designed to learn latent spatial covariance patterns for reconstructing sparse observations, we argue that these two reanalysis datasets are suitable for training the PCNN.

The GLORYS12v1 reanalysis[25], distributed by the Copernicus Marine Environment Monitoring Service (CMEMS), provides eddy-resolving global ocean/sea ice simulations at a horizontal resolution of 1/12° and 50 vertical layers. This system utilizes the NEMO3.1-LIM2 ocean-sea ice coupled model, assimilating satellite-derived SIC using a reduced-order Kalman filter. The C-GLORSv7[24] ocean-sea ice reanalysis, operated by the Euro-Mediterranean Center on Climate Change (CMCC), employs an eddy-permitting 0.25°×0.1° horizontal grid with 75 vertical levels. It builds upon the NEMO3.6-LIM2 ocean-sea ice coupled model, assimilating satellite-derived SIC through a 6-h nudging scheme.

Table 2. Details of ICESat's periods used in this study.

| ICESat period | Start date | End date |
|---|---|---|
| Laser 1 | 20 Feb 2003 | 29 Mar 2003 |
| Laser 2a | 25 Sep 2003 | 19 Nov 2003 |
| Laser 2b | 17 Feb 2004 | 21 Mar 2004 |
| Laser 2c | 18 May 2004 | 21 Jun 2004 |
| Laser 3a | 3 Oct 2004 | 8Nov 2004 |
| Laser 3b | 17 Feb 2005 | 24 Mar 2005 |
| Laser 3c | 20 May 2005 | 23 Jun 2005 |
| Laser 3d | 21 Oct 2005 | 24 Nov 2005 |
| Laser 3e | 22 Feb 2006 | 28 Mar 2006 |
| Laser 3f | 24 May 2006 | 26 Jun 2006 |
| Laser 3g | 25 Oct 2006 | 27 Nov 2006 |
| Laser 3h | 12 Mar 2007 | 14 Apr 2007 |
| Laser 3i | 2 Oct 2007 | 5 Nov 2007 |
| Laser 3j | 12 Feb 2008 | 21 Mar 2008 |
| Laser 3k | 4 Oct 2008 | 19 Oct 2008 |
| Laser 2d | 25 Nov 2008 | 17 Dec 2008 |
| Laser 2e | 9 Mar 2009 | 11 Apr 2009 |
| Laser 2f | 30 Sep 2009 | 11 Oct 2009 |

**Sea ice thickness products for validation and comparison**

We validated the PCNN-reconstructed SIT using data from upward-looking sonar (ULS) and two radar altimeter-derived products. The ULS dataset[46], maintained by the Alfred Wegener Institute (AWI), provides sea ice draft measurements collected from 13 fixed mooring locations in the Weddell Sea between 1990 and 2011, totalling over 3.7 million measurements with an uncertainty range of ±5 to ±23 cm. The SIT ($h$) was calculated from the corrected draft[47] ($d$) by $h = 0.028 + 1.012 * d$.

The European Space Agency's Sea Ice Climate Change Initiative project[48,49] (ESA SICCI) provides SIT estimates from Envisat and CryoSat2, covering the entire Southern Ocean since 2002. Although this monthly dataset has significant uncertainties[6], it remains an observational reference for hemispheric-scale model evaluations[31,32,50].

Another monthly SIT dataset is provided by the Laboratory of Space Geophysical and Oceanographic Studies[4] (LEGOS) and covers 30 years (1994–2023) for both polar regions. This product synthesizes data from four radar altimeters (ERS-1/2, Envisat, and CryoSat-2) through radar freeboard conversion, incorporating snow depth and density from multiple sources. Weighted averaging was applied to snow depth estimates to mitigate product uncertainties. The final SIT dataset was gridded at 12.5 km resolution and validated against in situ and independent satellite products.

It is noteworthy that SMOS thin ice observations were not included in the comparative analysis due to the unavailability of in-situ ULS measurements during the overlapping period between our reconstruction and SMOS observations.

**Data preparation and pre-processing**

The Equal-Area Scalable Earth Grid[51] (EASE-Grid 2.0), hereafter EASE, was used in this study to address polar projection distortions inherent in geographic coordinate grids. All of the aforementioned data were bilinearly remapped to the Southern Hemisphere 12.5 km EASE grids, which consist of 700 rows and 700 columns (totalling 490,000 grid points).

The equivalent SIT ($h_e$) was calculated on 12.5 km EASE grids through element-wise multiplication of observed actual SIT ($h$) from ICESat/ICESat-2 with SIC from AMSR-E/AMSR2 (Eq. 1) and will be used for subsequent processing and reconstruction:

$$h_e(t, x, y) = h(t, x, y) \times SIC(t, x, y), \qquad (1)$$

where $t$ denotes time and ($x$, $y$) represents the EASE grid coordinate. Similarly, the SIT from SICCI, LEGOS, and ULS is converted to equivalent SIT for comparative analysis.

To mitigate the impact of sparse daily satellite orbital sampling on reconstruction accuracy, a 5-day moving averaged SIT is calculated from ICESat and ICESat-2 data. Similarly, the daily SIT from reanalyses and SIC observations were processed into 5-day moving averaged fields to ensure temporal alignment.

To accelerate the training process and ensure consistency, SIT from reanalysis and satellite data are normalized using multi-year (1993–2020) monthly means and monthly standard deviations calculated from GLORYS12v1.

**Deep learning framework: from training to testing**

We implement a PCNN with a U-net architecture to reconstruct Antarctic SIT from sparse ICESat/ICESat-2 data (Fig. 1 & Table 3). The PCNN's input consists of training data concatenated with binary masks (1 for observed grid points, 0 otherwise) derived from satellite orbital coverage. During the encoding phase, partial convolutional operations progressively reduce the spatial dimensions of inputs while increasing the number of feature channels. This process extracts local spatial features and captures high-level features through stacked convolutional layers. In the decoding phase, the PCNN maps these high-level features back to the initial reanalysis data. Through iterative training, the PCNN learns to reconstruct the complete SIT distribution from limited observational data. Additionally, skip connections transmit input information to deeper layers for efficient reconstruction of Antarctic SIT.

Figure 2 illustrates the workflow encompassing data partitioning through spatially complete SIT field reconstruction. During the training phase, the reanalysis data were divided as follows: Daily SIT fields from C-GLORSv7 (1989–2011) and GLORYS12v1 (1993–2011), comprising 15,339 samples in total, were employed for training the PCNN model. Their 2012–2015 data (2,922 daily samples) was the validation set for hyperparameter tuning. The testing set contained 3,288 daily samples from GLORYS12v1 (2016–2020) and C-GLORSv7 (2016–2019), which were exclusively reserved for evaluating the PCNN's reconstruction performance.

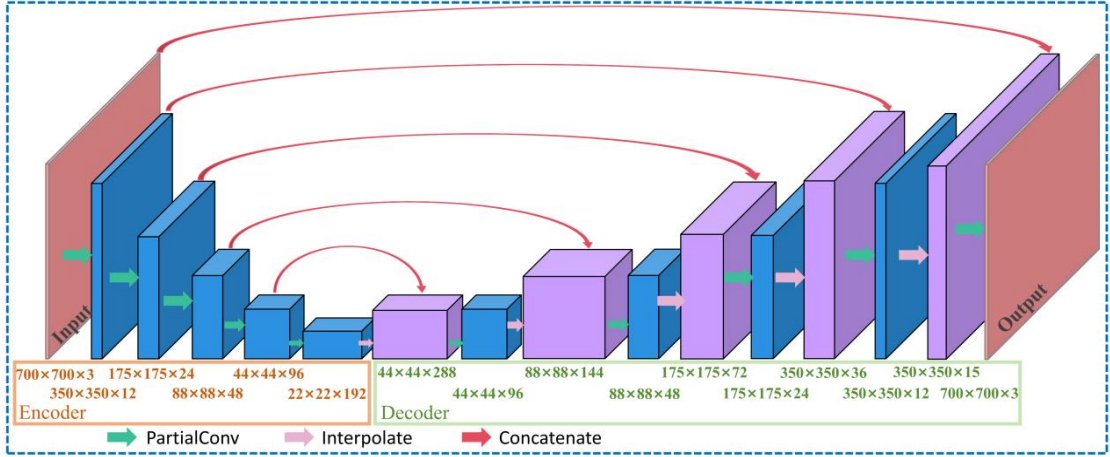

Fig. 1 Schematic diagram of structure for the PCNN model, which includes an encoding and a decoding part. The leftmost "box" is the input to the DLM and the rightmost "box" is the output, with the different operations (e.g., partial convolution, interpolation, and concatenation) indicated by arrows with different colors. The numbers below each "box" represent the shape of the output data from the various operations.

Table 3. Details of the PCNN architecture. Pconv, UpSample, Concat, and ReLU represent the partial convolutional layers, nearest neighbor interpolation, skip connections and rectified linear unit, respectively.

| Module name | Filter size | Channels | Stride | Padding | Nonlinearity |
|---|---|---|---|---|---|
| PConv1 | 7×7 | 12 | 2 | 3 | ReLU |
| PConv2 | 7×7 | 24 | 2 | 3 | ReLU |
| PConv3 | 5×5 | 48 | 2 | 2 | ReLU |
| PConv4 | 5×5 | 96 | 2 | 2 | ReLU |
| PConv5 | 3×3 | 192 | 2 | 1 | ReLU |
| UpSample1 | - | 192 | - | - | - |
| Concat1 | - | 288 | - | - | - |
| Pconv6 | 3×3 | 96 | 1 | 1 | LeakyReLU |
| UpSample2 | - | 96 | - | - | - |
| Concat2 | - | 144 | - | - | - |
| Pconv7 | 3×3 | 48 | 1 | 1 | LeakyReLU |
| UpSample3 | - | 48 | - | - | - |
| Concat3 | - | 72 | - | - | - |
| Pconv8 | 3×3 | 24 | 1 | 1 | LeakyReLU |
| UpSample4 | - | 24 | - | - | - |
| Concat4 | - | 36 | - | - | - |
| Pconv9 | 3×3 | 12 | 1 | 1 | LeakyReLU |
| UpSample5 | - | 12 | - | - | - |
| Concat5 | - | 18 | - | - | - |
| Pconv10 | 3×3 | 3 | 1 | 1 | LeakyReLU |

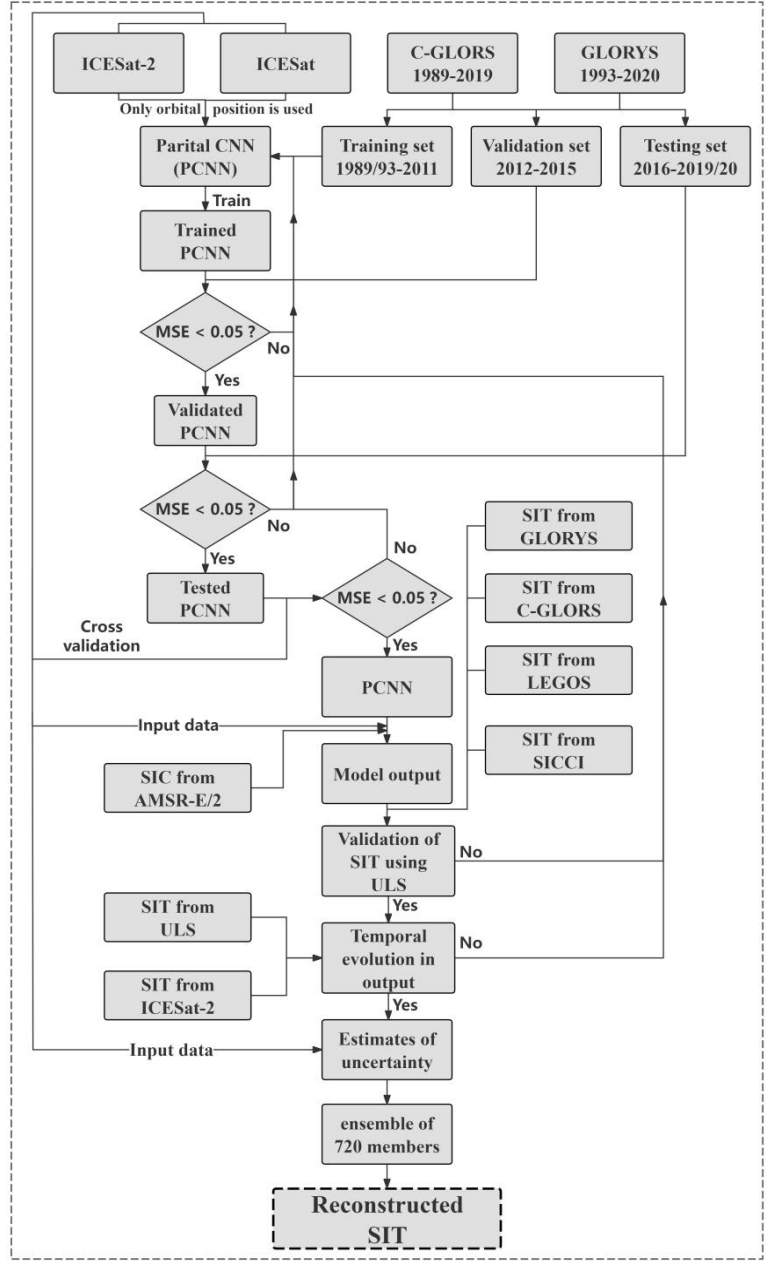

Fig. 2 Schematic view of Antarctic sea ice thickness (SIT) reconstruction. "MSE", "PCNN", "ULS", and "SIC" represent the mean squared error, partial convolutional neural network, upward-looking sonar, and sea ice concentration, respectively. "LEGOS" and "SICCI" refer to the two radar altimeter-derived Antarctic SIT datasets.

The mean squared error (MSE) between reconstructed and reanalysis SIT serves as the loss function, guiding parameter optimization through gradient descent. Training is terminated to prevent overfitting if the loss on the training set does not decrease for 20 iterations or decreases on the training set while increasing on the validation set. Ultimately, the DLM was trained for 11,000 iterations with a batch size of 50 and a learning rate of $2 \times 10^{-4}$.

Two daily samples from February 15, 2016 and September 15, 2016 were selected to demonstrate the PCNN's reconstruction performance on the testing set, as these dates respectively correspond to the annual minimum and maximum sea ice extents in the Antarctic region (Fig. 3). By comparing the SIT fields (Fig. 3b, e) reconstructed from the sparse observation locations (Fig. 3a, d) and the original reanalysis fields (treated as ground truth; Fig. 3c, f), we can see that they have satisfactory spatial consistency. The correlation coefficients (R) and RMSE are 0.93 and 0.17 m for September 15, 2019 (Fig. 3e, f), respectively, and those for February 15, 2019 (Fig. 3b, c) are 0.89 and 0.32 m, respectively. Across the entire testing set (3,288 samples), the R and RMSE were 0.89 and 0.21 m, respectively, and all values were statistically significant ($P < 0.05$). The high correlation and low RMSE demonstrate the PCNN's robust capability to reconstruct full-coverage Antarctic SIT from sparse along-track SIT observations.

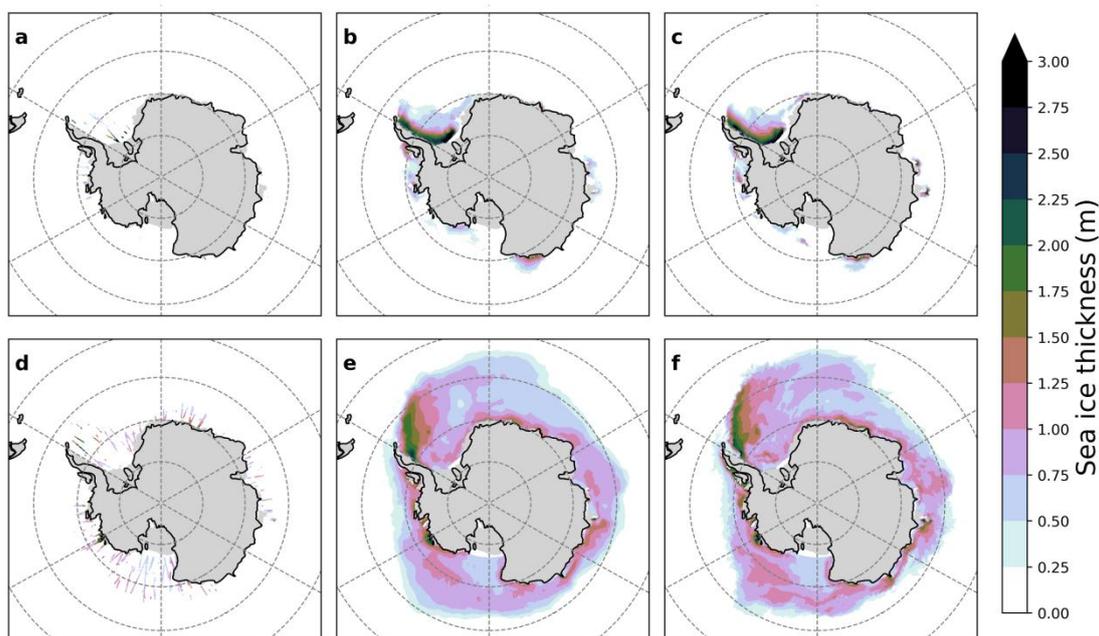

Fig. 3 Sea ice thickness patterns reconstructed on the testing set by the PCNN for (a-c) February 15 and (d-f) September 15, 2016. (a, d) Reanalysis data from the testing set, which is limited to locations with available observations. (b, e) SIT reconstructed by PCNN based on (a) and (d), respectively. (c, f) Original reanalysis data from the testing set, target for (b, e).

**Evaluation of PCNN via cross-validation**

In addition to the large-scale pattern validation, three cross-validation experiments were implemented to assess the PCNN's ability to resolve localized SIT features. The Southern Ocean was divided into six geographic sectors (Fig. 4): Bellingshausen and Amundsen Sea (B-A; 60°W-130°W), Western Weddell Sea (W-W; 45°W-60°W), Eastern Weddell Sea (E-W; 45°W-20°E), Indian Ocean (I-O; 20°E-90°E), Pacific Ocean (P-O; 90°E-160°E), and the Ross Sea (160°E-130°W).

In each cross-validation experiment, ICESat/ICESat-2 SIT observations were partitioned into two mutually exclusive subsets: one used for reconstruction, while the other subset composed of the data in the designated boxes was reserved as ground truth for accuracy assessment (Fig. 4). These validation boxes were distributed across distinct geographic sectors to isolate region-specific model performance variations. This box configuration was maintained for both ICESat (2003–2009) and ICESat-2 (2018–2024) periods, establishing a temporally consistent validation framework to evaluate model consistency across different sea ice regimes.

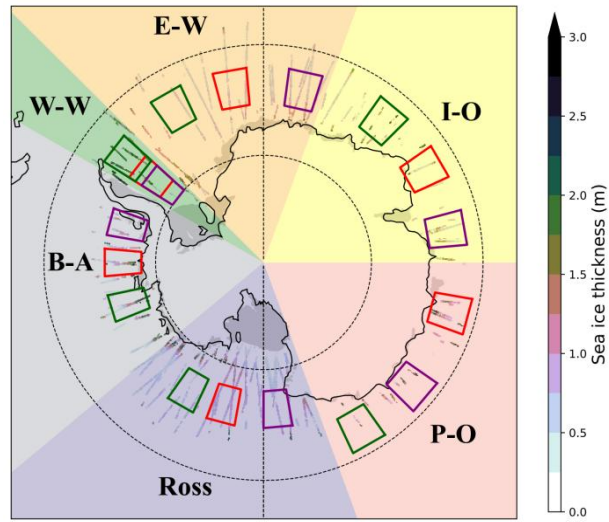

Fig. 4 Schematic diagrams of the "box" of the different geographic sectors used for cross-validation. The red, dark green, and purple "boxes" represent cross-validation experiments #1, #2, and #3. The tracks represent the SIT from satellite data on a randomly selected 5-day. The abbreviations "B-A," "W-W," "E-W," "I-O," "P-O," and "Ross" refer to specific sea regions: the Bellingshausen and Amundsen Seas, the Western Weddell Sea, the Eastern Weddell Sea, the Indian Ocean sector, the Pacific Ocean sector, and the Ross Sea, respectively.

The PCNN demonstrates high consistency in SIT reconstruction between the ICESat-2 (Fig. 5a-c) and the ICESat (Fig. 5d-f). Small RMSE between the reconstructed SIT and in-box observations are obtained in the W-W and Ross Seas, while large RMSE is in the P-O and I-O sectors. The evaluation results align well across different months, with higher RMSE during warmer periods, such as February–March and November–December (Fig. 5a, d). However, some discrepancies exist between the ICESat and ICESat-2 evaluations. For example, in cross-validation #2 for the E-W sector in September, the RMSEs are 1.1 (Fig. 5e; ICESat) and 0.23 (Fig. 5b; ICESat-2), respectively.

To further illustrate the potential factors affecting PCNN's reconstruction performance, we fitted the RMSE between reconstructed and in-box SIT observations (Fig. 6a-c) using a multivariate regression approach. The factors include the local SIT variability (Fig. 6d-f) and the number of observations (Fig. 6g-i) in each geographic sector and each month. The local SIT variability is quantified by standard deviation (STD). The overall results from the 50 regressions indicate that the STD (regression coefficient β = 0.907) and the observation number (β = −0.227) play an important role in SIT reconstruction, while sector and month show negligible effects (β < 0.01). The regression model achieves an explained variance of 0.8 and a mean absolute error (MAE) of 0.09 m, confirming that the local SIT spatial variability is the most influential factor, followed by the observation number.

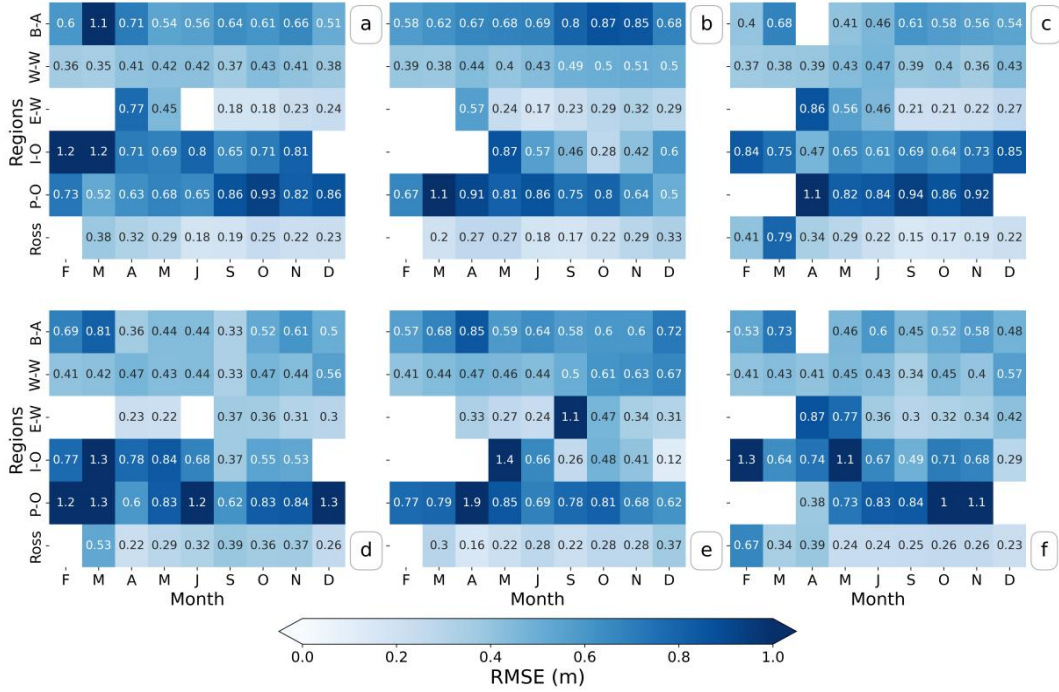

Fig. 5 Evaluation of SIT reconstructed by out-of-box observations for different months and sea regions using in-box data. The (a, d), (b, e), and (c, f) indicate cross-validations #1, #2, and #3, respectively. (a-c) Cross-validation during the ICESat-2 period, while (d-f) corresponds to the ICESat period. Only the months common to both the ICESat and ICESat-2 are analyzed. RMSE denotes Root Mean Square Error in meters. The abbreviations "B-A", "W-W", "E-W", "I-O", "P-O", and "Ross" are the same as in Fig. 4. The label "J" on the x-axis denotes June.

The PCNN reconstructs pan-Antarctic SIT by integrating satellite observations with learned spatial patterns of SIT from training data. To evaluate the added value of satellite data for SIT reconstruction, we validated the results against GLORYS12v1 and C-GLORSv7 reanalysis products utilizing independent ICESat-2 data within each cross-validation "box" (Fig. 7). Both reconstructed SIT (Fig. 5d-f) and reanalyses (Fig. 7a-c) exhibit higher RMSE in the B-A, P-O, and I-O sectors than in the W-W, E-W and Ross Seas, with reanalyses' RMSE further amplified during December–March (Fig. 7). Notably, the RMSE of the reconstructed SIT relative to ICESat-2 observations is consistently lower than that of reanalyses in most geographic sectors and months (Fig. 7d-f). The reconstruction demonstrates the most significant improvements in the B-A, P-O, and I-O sectors, where reanalyses exhibit the largest RMSE, particularly during December–March (excluding February due to insufficient sample sizes). These results highlight the urgent need to refine SIT estimates in these regions to improve our understanding of sea ice-related processes.

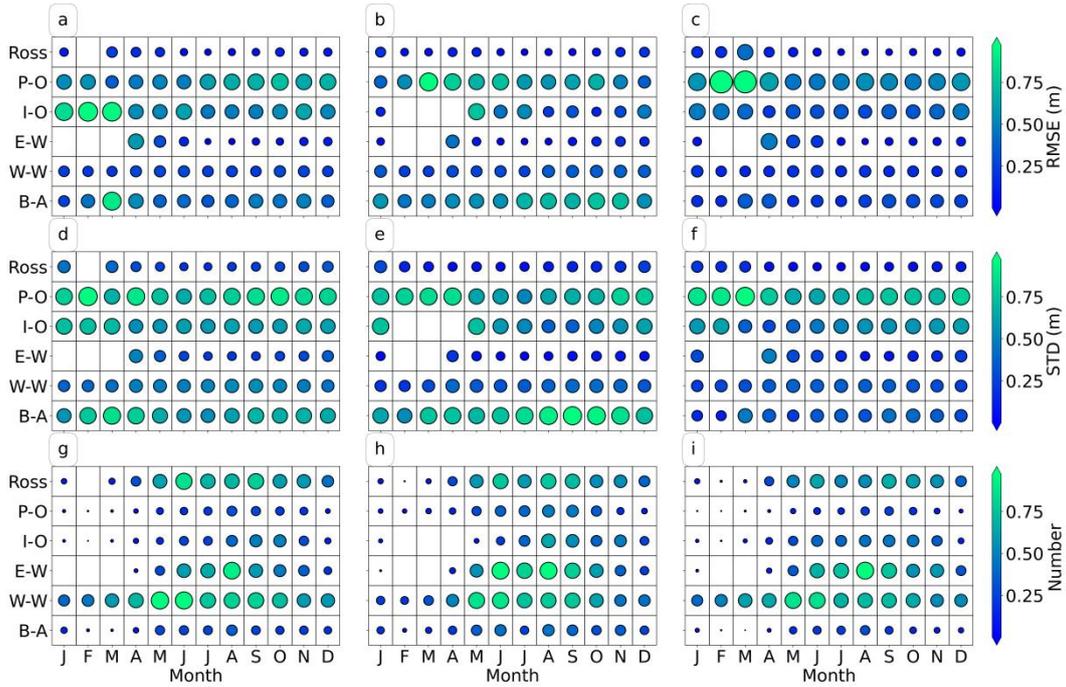

Fig. 6 The RMSE and statistics for different cross-validation. (a-c) RMSE between in-box ICESat-2 observations and the SIT reconstructed from out-of-box data. (d-f) SIT variability (standard deviation; STD) and (g-i) number of observations, respectively, for regions centered on the box shown in Fig. 4, expanded to four times the area. The number of observations was normalized using min-max scaling. Larger values are represented by larger circles with colors closer to green. The abbreviations "B-A," "W-W," "E-W," "I-O," "P-O," and "Ross" are the same as in Fig. 4.

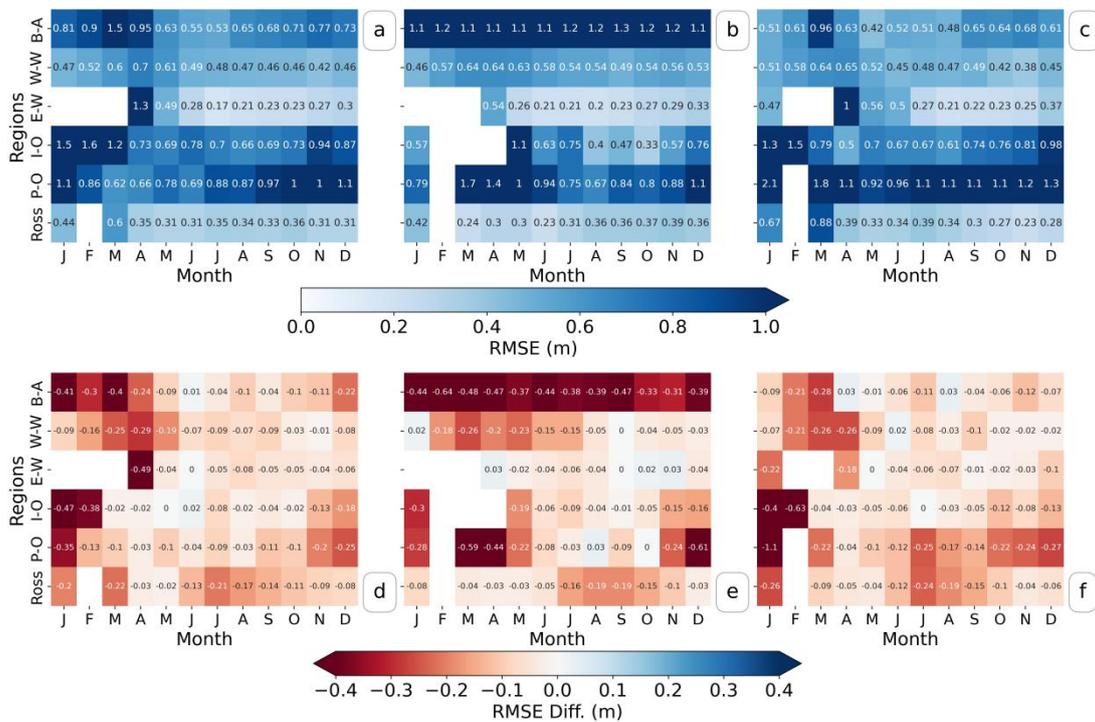

Fig. 7 Evaluation of the reconstructed SIT using reanalysis data and in-box observations. (a-c) RMSE between reanalysis and satellite in-box data for each cross-validation. (d-f) RMSE difference between reconstructed SIT versus reanalysis SIT and reanalysis SIT versus observed SIT for each cross-validation.

**Uncertainty estimates**

The uncertainty in the reconstructed SIT stems from three sources: (a) Along-track observation itself, which is influenced by sea ice/snow density uncertainties, total freeboard uncertainties, and empirical parameters[19]. (b) The training data variability. Considering that different training dataset periods may lead to variations in the SIT spatial relationships. (c) PCNN model variability, arising from stochasticity in mini-batch gradient descent training.

We estimate the total uncertainty at the EASE grid-point level as follows:

$$\sigma(t,x,y)^2 = \frac{1}{N}\left(\sum_{i=1}^{N} h_i(t,x,y)^2\right) - \mu(t,x,y)^2, \tag{2}$$

where $\sigma$ represents the total uncertainty, N = 720 is the number of ensemble members, $h_i$ is $i$-th reconstructed SITs, $\mu$ is the ensemble mean. The 720 ensemble combines 20 scenarios for along-track SIT uncertainty, 12 training datasets from different reanalysis periods (Table 4), and 3 independent PCNN trainings (each generating a distinct model).

Table 4. Division of the reanalysis data for estimating the uncertainty in the reconstructed SIT.

| Reanalysis datasets | Period | Application |
|---|---|---|
| GLORYS12v1 | 1993–2009 | Training |
| | 1995–2011 | |
| | 1997–2013 | |
| | 1999–2015 | |
| | 2001–2017 | |
| | 2018–2020 | Testing |
| C-GLORSv7 | 1989–2005 | Training |
| | 1991–2007 | |
| | 1993–2009 | |
| | 1995–2011 | |
| | 1997–2013 | |
| | 1999–2015 | |
| | 2001–2017 | |
| | 2018–2019 | Testing |

The reconstructed SIT exhibits spatially heterogeneous uncertainty: highest in the ice edge and coastal ice zones, and lowest in the central ice zone (Fig. 8). In the ice edge zone, rapid ice growth/melt and sparser satellite coverage present a greater challenge for accurately reconstructing SIT, especially during seasonal transition periods. In the coastal ice zones, SIT reconstruction is subject to greater uncertainty due to the exclusion of the freeboard within 25 km of land. Seasonally, uncertainty is higher in ice-melting months (October–March) than in ice-freezing months (April–September; Fig. 8 & 9), with no significant inter-annual trend (2018–2024; Fig. 9). Among all geographic sectors, the P-O sector exhibits the highest uncertainty across all seasons, attributed to its higher SIT spatial pattern and sparser observations (Fig. 9). The B-A and I-O sector follow with intermediate uncertainty during ice-melting months (October–March), while the W-W sea ranks similarly levels during ice-freezing months. In contrast, the E-W and Ross sectors typically show lower uncertainty, with the Ross Sea remaining the most stable region year-round (Fig. 9). Uncertainty estimates were unavailable for the colder periods of 2019 and 2024 due to the absence of along-track observations.

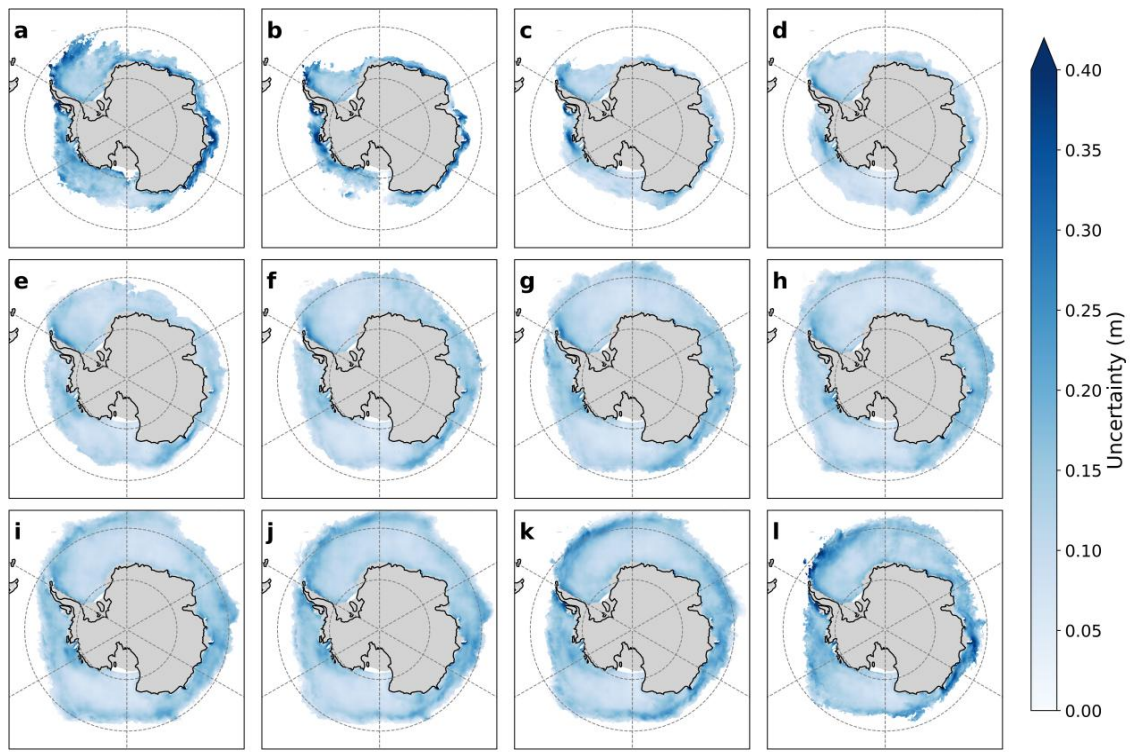

Fig. 8 Uncertainty in the reconstructed SIT across different months from October 2018 to August 2024. (a-l) The SIT uncertainty from January to December, respectively.

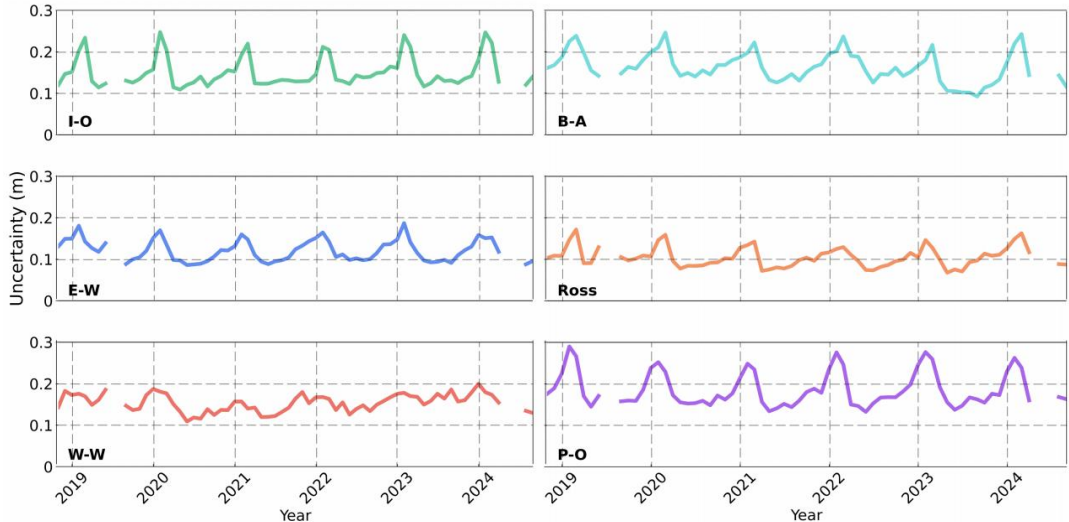

Fig. 9 Uncertainty in reconstructed SIT across different geographic sectors from October 2018 to August 2024. The abbreviations "B-A," "W-W," "E-W," "I-O," "P-O," and "Ross" are the same as in Fig. 4.

## Data Records

The dataset is available at Figshare[52]. This study reconstructs Antarctic SIT using observations from ICESat and ICESat-2. The dataset includes reconstructed SIT from ICESat-2 starting from October 2018, as well as a reconstruction based on ICESat from 2003 to 2009, which contains temporal discontinuities (see Table 2). The uncertainty of the reconstructed SIT is estimated using 720 ensemble members, as detailed in Section *Uncertainty estimates*. The reconstructed results are stored in NetCDF format files with the EASE grids. In these files, "thick" represents the ensemble-mean SIT, while "error" denotes the uncertainty in reconstructed SIT. "x," "y," and "time" correspond to the row and column of the EASE grids and the time variable, respectively, while "latitude" and "longitude" represent the geographic coordinates. This dataset can be continuously updated using the code provided in Section *Code Availability*.

## Technical Validation

### Validation of reconstructed sea ice thickness using ULS data

Figure 10 shows the validation of the reconstructed results from ICESat using independent ULS SIT observations, which are unavailable during the ICESat-2 period. The details of the ULS observations' locations and periods are listed in Table 5. The ocean reanalyses used for training (GLORYS12v1 and C-GLORSv7) and two other satellite-derived SIT products (LEGOS and SICCI) are also compared. To ensure temporal consistency with the monthly products LEGOS and SICCI, the reconstructed 5-day resolution SIT was processed for monthly averaging. Among the total 23 validation cases, 19 (83%) cases show that the uncertainty range of the reconstructions covers the ULS observations (Fig. 10), while only 4 cases (Fig. 10c, e, m, n) exhibit minor deviations. The reconstructed SIT has an MAE of 0.238m compared to ULS across 23 validations, outperforming GLORYS12v1 (0.297 m), C-GLORSv7 (0.371 m), LEGOS (0.543 m), and SICCI (1.276 m). Additionally, the reconstruction uncertainty is significantly lower than that of LEGOS and SICCI. GLORYS12v1 shows better agreement with ULS than C-GLORSv7, LEGOS, and SICCI; in 15 of 23 cases, it falls within the uncertainty range of the reconstructed SIT.

It is important to note that the cross-validation experiments demonstrate that the PCNN-reconstructed SIT exhibits consistent performance across various missions and remains temporally stable. This evidence allows the validation results established with ULS during the ICESat era (2003–2009) to be reliably extended to the ICESat-2 period (2018–2024).

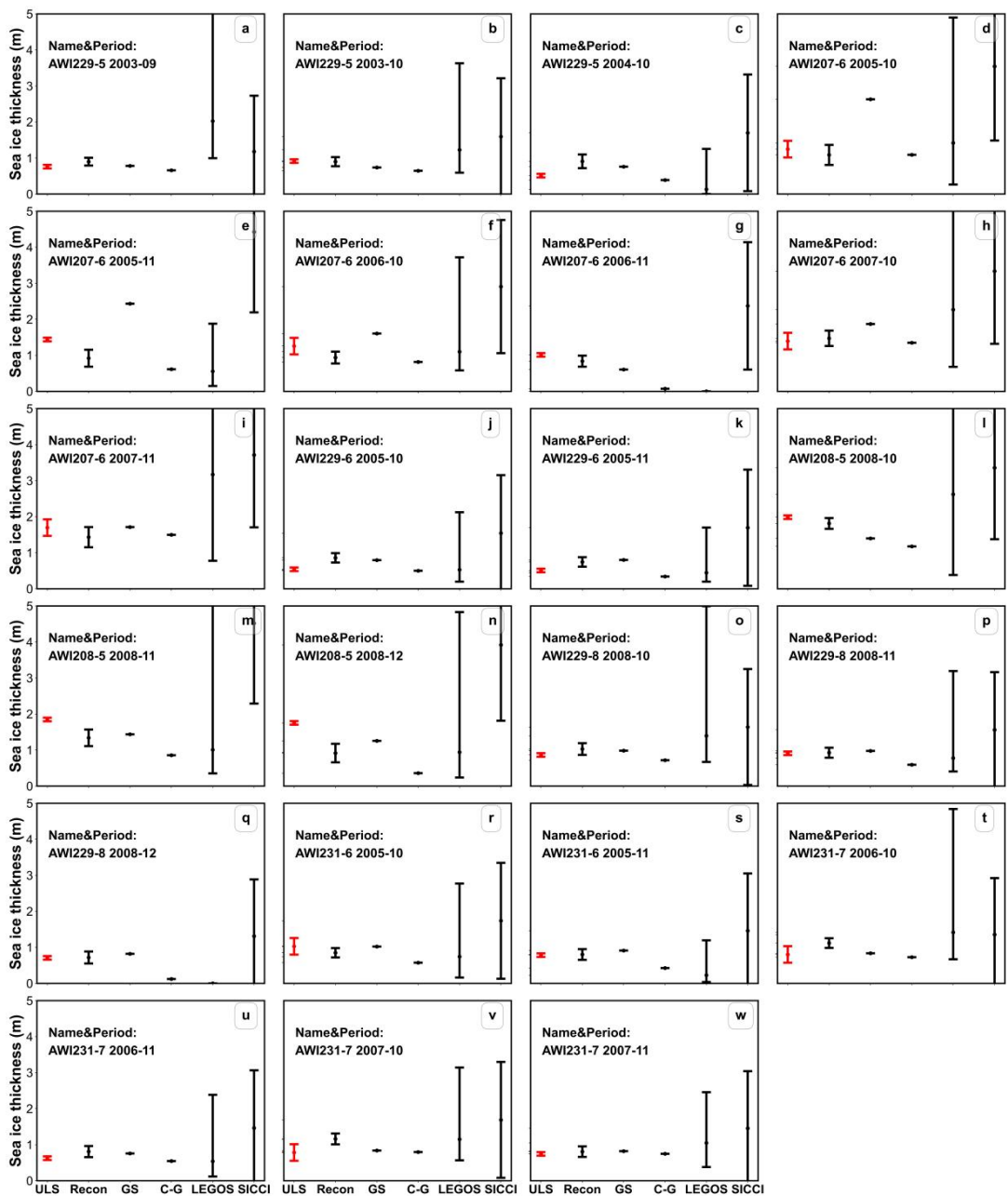

Fig. 10 Comparison of the PCNN-reconstructed SIT, GLORYS12v1, C-GLORSv7, LEGOS, and SICCI with observations from the ULS on a monthly scale. The IDs and periods of the ULS are listed, and the location of the ULS refers to Table 5. "Recon," "GS," and "C-G" denote the reconstructed SIT, as well as SIT from GLORYS12v1 and C-GLORSv7, respectively. Error bars represent the uncertainty, which is not available for the reanalyses. Uncertainty estimates for the reconstructed SIT are detailed in Section *Uncertainty estimates*.

Table 5. Details of the ULS used to evaluate the temporal evolution of the reconstructed SIT.

| ULS Name | Latitude | Longitude | Period |
| --- | --- | --- | --- |
| AWI206 | -63.480 | -52.096 | 199605–199801 & 200804–201101 |
| AWI207 | -63.703 | -50.870 | 199011–199211 & 199605–199711 & 200503–200809 |
| AWI208 | -65.614 | -37.407 | 199301–199407 & 200803–201101 |
| AWI210 | -69.660 | -15.715 | 199012–199212 |
| AWI212 | -70.912 | -11.963 | 199012–199212 |
| AWI217 | -64.418 | -45.850 | 199011–199211 |
| AWI227 | -59.029 | -0.015 | 199601–200012 |
| AWI229 | -63.954 | -0.004 | 199604–200512 & 200802–201012 |
| AWI230 | -66.006 | 0.173 | 199901–200212 |
| AWI231 | -66.511 | -0.032 | 199604–199907 & 200012–200212 & 200502–200803 |
| AWI232 | -68.998 | -0.005 | 199604–199702 & 199901–200502 & 200512–201012 |
| AWI233 | -69.394 | -0.066 | 199701–199803 & 200212–200502 |

**Validation of reconstructed SIT's temporal evolution**

Given that the PCNN primarily reconstructs SIT fields by learning spatial relationships of SIT from the training data, the physical plausibility of its temporal evolution requires verification. The monthly climatologies derived from ULS observations are adopted as the reference. Figure 11 reveals strong agreement between the reconstructed SIT and ULS SIT, with statistically significant high correlations (R>0.9) observed at 9 of the 12 validation sites (Fig. 11). Importantly, the reconstructed SIT captures the seasonal growth and melt, characterized by sea ice thickening from February–March and thinning from September–October (Fig. 11). However, the correlation at AWI212 (Fig. 11e), AWI217 (Fig. 11f) and AWI233 (Fig. 11-l) exhibit relatively low, possibly due to the inherent limitations of the 12.5 km grid-averaged SIT in representing the subgrid-scale SIT features measured by ULS. Caution should be exercised when comparing absolute SIT values, as the temporal coverage of the ULS (1990–2011) and reconstructed SIT (2018–2024) do not overlap.

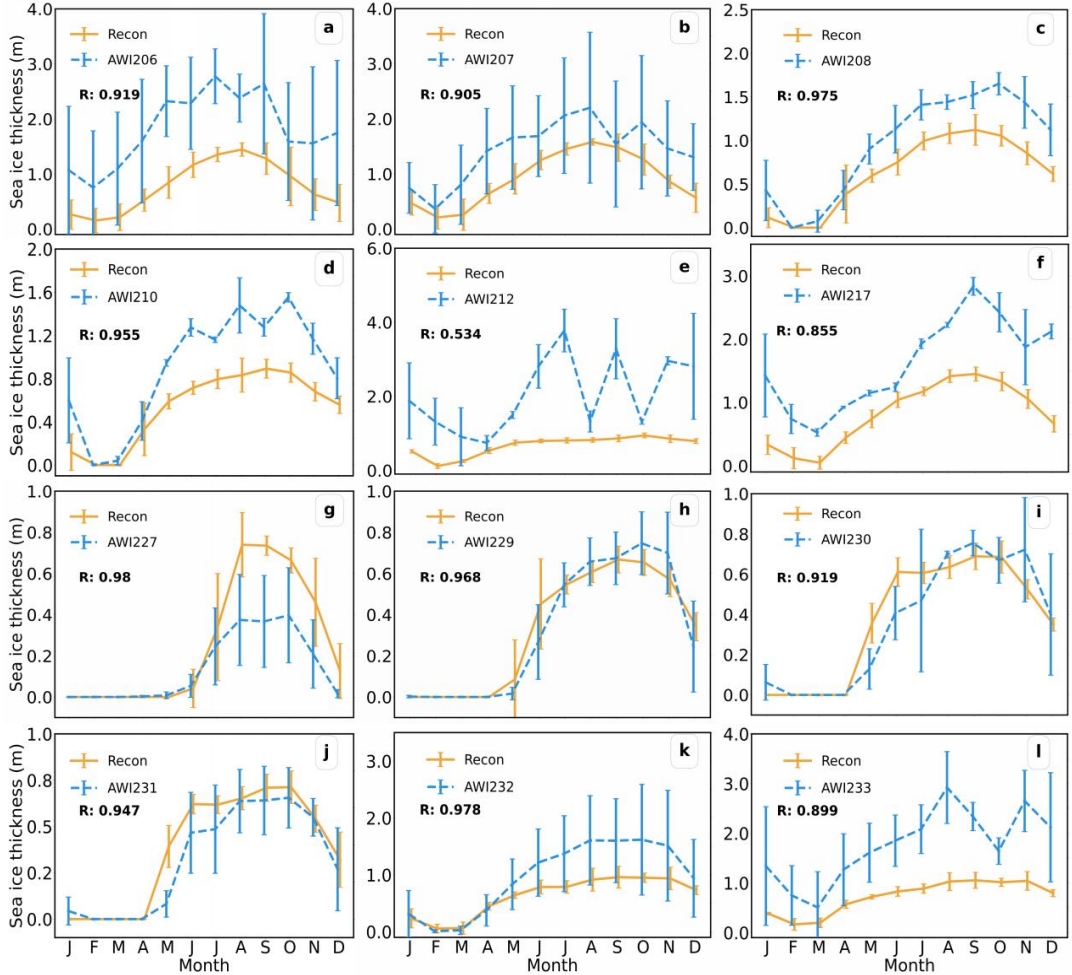

Fig. 11 Comparison of the climatology of reconstructed SIT with ULS observations. The orange line represents the monthly climatology of the reconstructed SIT from October 2018 to August 2024, while the blue line represents the ULS climatology (periods are listed in Table 5). The error bars indicate the inter-annual variability. "R" denotes the correlation coefficient between the blue and orange lines.

We quantified the temporal tendency at grid-cell resolution to validate the reconstructed SIT's intra-seasonal evolution further. This analysis is grounded in the fundamental thermodynamic principle that SIT generally increases during austral autumn/winter and decreases in spring/summer. The tendency of SIT was computed following Eq. (3):

$$\frac{\partial h}{\partial t} = \frac{1}{2(N-2)} \sum_{n=2}^{N-1} (h(x,y)^{n+1} - h(x,y)^{n-1}) \qquad (3)$$

where $h$ denotes SIT, $N$ is the number of days in the season, and $(x, y)$ are the EASE grids. The reconstructed SIT basically follows the thermodynamic principle (Fig. 12). However, there exists three regions (marked as A, B, and C) where SIT is significantly thinning in winter (Fig. 12b). To validate this unexpected winter thinning, ICESat-2 along-track observations in these three regions are utilized to calculate the winter regional mean SIT and its temporal evolution using Eq. 3 (Fig. 13).

In regions A and B, wintertime SIT exhibited gradual increases but showed an abrupt reduction during early winter, resulting in an overall negative trend over the winter season (−0.299 cm/day in A and −0.337 cm/day in B). Region C displayed higher SIT variability than regions A and B, with progressive thickening interrupted by late-winter episodic thinning events resulting in a net negative trend (−0.162 cm/day). These results confirm that the intra-seasonal variability in the reconstructed SIT fields predominantly originates from ICESat/ICESat-2 observational constraints, thereby validating their physical plausibility. Notably, all three regions are frequent polynya formation areas[53], where sudden SIT reductions may stem from oceanic warm water upwelling (regions A/B) and enhanced offshore wind-driven divergence (region C).

While the reconstructed SIT captures seasonal and intra-seasonal variability well, its representation of day-to-day variations is limited by the sparse and uneven distribution of along-track satellite observations. The 5-day moving window introduces temporal inconsistencies, as daily estimates are influenced by observations from adjacent days. Additionally, ice drift and sparse sampling can cause abrupt and possibly spurious jumps in SIT. While thick ice can be detected, its spatial extent remains uncertain due to limited observational coverage. As a result, users are advised to interpret daily SIT variations with caution, particularly in studies focused on short-term dynamics or regional processes. Future improvements should aim to integrate observations from multiple satellite missions to reduce sampling biases.

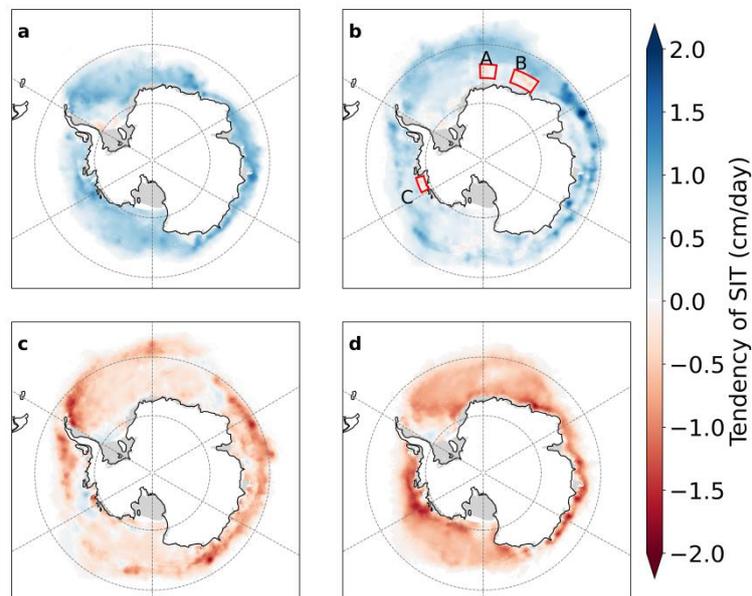

Fig. 12 The seasonal evolution of reconstructed SIT from October 2018 to August 2024. (a-d) The reconstructed SIT changes during autumn, winter, spring, and summer, respectively. Blue indicates increases in SIT, while red denotes decreases.

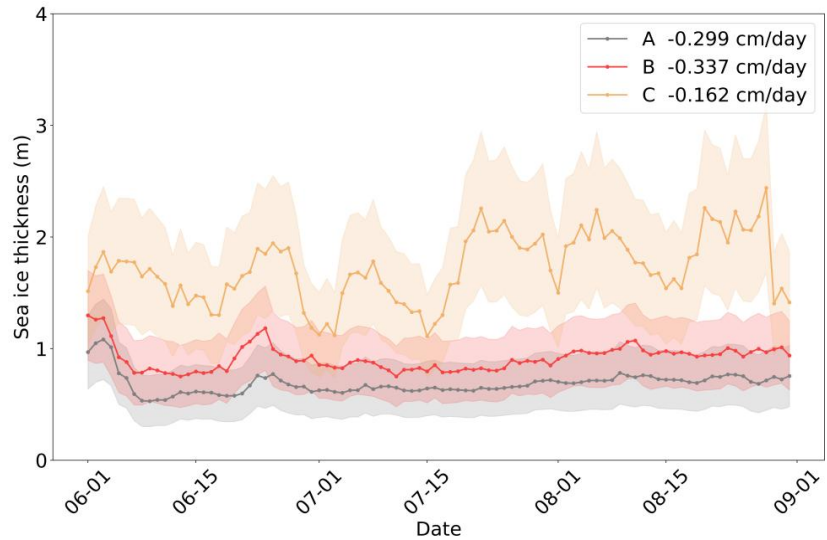

Fig. 13 The seasonal evolution of the regionally averaged SIT from ICESat-2 data during winter in the three regions A, B, and C shown in Fig. 12b. The shaded areas represent SIT's uncertainty.

## Code Availability

The code used in this study can be found at Figshare[54]. This code may be updated over time.

## Acknowledgements

This work was supported by the Southern Marine Science and Engineering Guangdong Laboratory (Zhuhai) (SML2022SP401, 2023SP219), the National Natural Science Foundation of China (No. 42406252), and the China Postdoctoral Science Foundation (No. 2023M741526). PU is funded by the EU Horizon 2020 PolarRES project (grant number 101003590) and by the Finnish Research Council (grant number 364876). We express our sincere appreciation to the authors and institutions that contributed freeboard observations from ICESat-2, as well as the providers of SIC data from AMSR-E and AMSR2. Our gratitude is also extended to the providers of the reanalysis datasets GLORYS12v1 and C-GLORSv7 for their invaluable data contributions. Special gratitude is given to Dr. Stefan Kern, who provided the total freeboard from the ICESat sea ice altimetry data. Furthermore, we thank Chao Min and Yifan Wang from the School of Atmospheric Sciences, Sun Yat-sen University, and Southern Marine Science and Engineering Guangdong Laboratory (Zhuhai) for helpful discussion.

## Author contributions

Q.Y., Y.N., and Z.M. designed the research and analysed the results; Y.X., D. X., and Q.S. collected data; Z.M. and W.S. verified the approach and conducted computations with the help of H.L.; Z.M., Y.N., Q.Y., and P.U. drafted. All other authors contributed to improving the manuscript.

## Competing interests

The authors declare no competing interest.